\def\a{\alpha}

\def\d{\delta}

\def\s{\sigma}

\def\D{\Delta}

\def\pvl{{\it Phys. Rev. Lett.}}

\def\a{\alpha}

\def\D{\Delta}

\def\IR{\relax{\rm I\kern-.18em R}}
\font\cmss=cmss10 \font\cmsss=cmss10 at 7pt
\def\IZ{\relax\ifmmode\mathchoice
{\hbox{\cmss Z\kern-.4em Z}}{\hbox{\cmss Z\kern-.4em Z}}
{\lower.9pt\hbox{\cmsss Z\kern-.4em Z}}
{\lower1.2pt\hbox{\cmsss Z\kern-.4em Z}}\else{\cmss Z\kern-.4em Z}\fi}
\def\IN{\relax{\rm I\kern-.18em N}}

\documentstyle[preprint,aps]{revtex}

\newcommand{\bge}{\begin{equation}}
\newcommand{\ege}{\end{equation}}
\newcommand{\bga}{\begin{eqnarray}} 
\newcommand{\ega}{\end{eqnarray}}

\begin{document}
\draft
%\newcomman{\baselinestretch}{1.4in}
%\baslineskip 1in
\topmargin -.5in
\textheight 9in
\textwidth 6.5in
\title{Mutual influence of structural distortion and superconductivity in
systems with degenerate bands}
\author{Haranath Ghosh and S. N. Behera}
\address{Institute of Physics, Bhubaneswar -751 005, INDIA.}
\author{S. K. Ghatak}
\address{Department of Physics and Meteorology, IIT, Kharagpur, INDIA.}
\author{and D. K. Ray}
\address{Laboratoire, P M T M, CNRS, Universite Paris-Nord, 93430,
Villetaneuse, France.} 
\maketitle
\begin{abstract}
The interplay between the band Jahn-Teller distortion and the
superconductivity is studied  for the system whose Fermi level
lies in two-fold degenerate band. Assuming that the lattice distortion is
coupled to the orbital electron density and the superconductivity 
arises due to BCS pairing mechanism between the electrons, the phase
diagram is obtained for different doping with respect to half-filled
band situation. The coexistence phase of superconductivity and distortion
occurs within limited range of doping and the distortion lowers 
the superconducting transition temperature $T_c$. In presence of strong
electron-lattice interaction the lattice strain is found to be maximum at
half-filling and superconductivity does not appear for
low doping. The maximum value of $T_c$ obtainable for an optimum doping
is limited by the structural transition temperature $T_s$. The growth
of distortion is arrested with the onset of superconductivity and the 
distortion is found to disappear at lower temperature for some hole
density. Such arresting of the growth of distortion at $T_c$ produces 
discontinuous jump in thermal expansion coefficient. The variation of
strain with temperature as well as with doping, thermal expansion 
coefficient, the $T_c$ vs $\delta$ behaviour are in qualitative agreement
with recent experimental 
observations on interplay of distortion and superconductivity
in cuprates.
\end{abstract}
%\maketitle
\section{Introduction}
The role of structural instability on the superconductivity
had been a crucial question since the discovery of high temperature
superconductivity \cite{1}. Apart from the intermetallic compounds
\cite{2}, the structural transition associated with the lowering of
crystallographic symmetry has also been observed in high-$T_c$ cuprates
\cite{3,4} and fullerenes \cite{5}. The existence of orbital degeneracy
with the Fermi level (FL) lying within a degenerate band is a common
feature to all these systems. Now for almost any set of degenerate
electronic states associated with a molecular configuration there will
exist some symmetry breaking interaction in which molecular distortion is
associated with the removal of electronic degeneracy \--- goes by the name
of Jahn--Teller theorem. Associated with such transition there is a change in
the electronic density of states (DOS) around the FL and therefore, it is
expected that the band Jahn-Teller (BJT) distortion would strongly
influence the superconductivity in such a system. If the FL lies at the
centre of the degenerate band then the DOS at the FL will be reduced
drastically and the superconducting (SC) transition temperature $T_c$ will
be reduced. On the other hand, $T_c$ can also be higher as phonon softening due
to the BJT effect may increase the coupling constant which is inversely
proportional to mean square phonon energy.
\par The dependence of $T_c$ on structural properties is less
understood. There is a large number of growing experimental evidence 
indicating the strong mutual influence of structural distortion and 
superconducting transition. In fact, Bednorz and M\"{u}ller in their
historic paper conjectured that a J-T type of polaron formation may lead
to high-$T_c$. Indeed, experiments indicate that in $(La_{1-x}M_x)_2CuO_4$
(M = Ca, Ba, Sr) with the onset of $T_c$ around 38K each $CuO_6$
octahedron is distorted with interlayer $Cu-O$ distance equal to
$1.9\AA$ and an elongated $Cu-O$ distance of $2.4\AA$. A detailed neutron
scattering measurement of the temperature dependence of the spontaneous
strain has been reported by Mck Paul et al, \cite{6} for
$La_{1.85}Ba_{0.85}CuO_4$ ($T_c~\approx 38K$). The system exhibits a
structural transition at 180 K ; the magnitude of the strain rises on
lowering temperature and shows an anamolous suppression below $75K$.
Although the suppression of the lattice distortion appears at a
temperature higher than $T_c$, it is expected that this anamoly is
associated with the appearance of superconductivity. Such structural
transition has also been observed in other high $T_c$ compounds and its
strong influence being correlated with the superconductivity by many
authors \cite{7}. The evidence for such interplay between lattice
distortion and superconductivity in $La_{1.85}Sr_{0.85}CuO_4$,
$YBa_2Cu_3O_{6.5}$ and even in the electron doped cuprate system
$Nd_{1.85}Ce_{0.15}CuO_{4-\delta}$ has been reported by Lang et al,
\cite{8} from thermal expansion measurements. These measurements while
demonstrating the insensitivity of structural distortion to charge
carriers doping clearly exhibits that the highest $T_c$ is limited by
lattice instability. Furthermore, the suppression of lattice distortion in
the SC state has also been seen from extended absorption of X-ray fine
structure edge measurements in the other families of cuprates e.g, the
`2212' and `2223' Bi-superconductors \cite{9}.
\par The structural transition from tetragonal to orthorhombic phase in
$La_{2-x}Sr_xCuO_4$ takes place at a temperature higher than the superconducting
transition temperature. The transition leads to a change of bond distance
in $CuO_2$ plane and the orthorhombic strain $\frac{a-b}{a+b}$ where $a$
and $b$ are bond lengths in the $CuO_2$ plane is the measure of
distortion. Recently this distortion is measured in $La_{2-x}M_xCuO_4$ (M
= Ca, Sr) as a function of doping ($x$). The structural transition is
found to decrease with $x$ and also it causes suppression of $T_c$. The
distortion decreases with $x$ in the superconducting phase and is
completely suppressed at a critical value of $x$ \cite{10}. These
measurements are clear demonstration of strong interplay between
superconductivity and structural distortion \cite{10,11}.
\par In order to have a qualitative understanding of the interplay
between the SC state and distortion, we present in this
communication a detailed study of the coexistence of
superconductivity and the BJT distortion for systems having FL lying
in a two fold degenerate
$e_g$-band, as a function of hole concentration. We find 
that, many unusual experimental findings like
spontaneous suppression of strain close to
$T_c$, kink like structures in the thermal expansion coefficient,
$T_c~ \sim \delta$ behavior and the role of ionic size of the
substituent ion on phase diagram etc. follows from this model. In
the next section we present the model 
calculation for our results briefing the essential ingredients of our
model and its inputs. Section 3 is devoted to analyze the calculated
results so that a qualitative comparison can be made with the observed
one. Finally, we conclude in section 4.
\section{Model $\&$ Calculation}

It is generally believed that in high temperature superconductors 
the electrons in $CuO_2$ layers
contribute most to normal and superconducting properties.
Experimental evidences also suggest that these charge carriers
occupy mainly the $p-state$ of oxygen. These electrons are also
coupled to lattice as the site energy would depend on the state
of the $CuO_2$ layer. In the tetragonal phase, electrons will `see'
the same site energy in $`a$' and $`b$' directions. However, 
orthorhombic distortion will remove such degeneracy of site energy as
the ligand field would shift the energy level differently in $a$ and $b$
direction of a unit cell. Such distortion will alter the electronic states,
 energy and thereby will influence superconductivity. A realistic
description of $CuO_2$ layer needs a five states \cite{12} model study to
examine interplay between superconductivity and lattice distortion, becomes 
an involved one. Instead we study a simple and tractable model that 
incorporates essential effect (removal of orbital degeneracy) of distortion 
and superconductivity. We consider a model system where a two-fold 
degenerate ($e_g)$ orbital is associated with each lattice point.
The orbitals are strongly coupled to lattice and forms a band. This
Jahn-Teller Hamiltonian can be described by,
\begin{equation}
H=\sum_{k,\sigma,\alpha=1,2} (\epsilon_k - \mu) C_{k
\alpha\sigma}^\dagger C_{k\alpha\sigma} + \sum_{k,\sigma} G e
(\hat n_{k1\sigma} - \hat n_{k2\sigma}) + \frac{1}{2} c e^2
\end{equation}
The first term represents band energy of electrons in two fold
degenerate band with $\mu$ being the chemical potential. The second
term describes the electron - lattice ligand potential due to
distortion $e$ which is coupled to population density of
orbitals ($n_{k\alpha}$). The expansion in direction $`a$'
lifts the one-electron level of orbital (say -1) whereas that of
label (say-2) goes down due to compression along $b$-direction.
In the limit of zero-volume change such shift of levels is equal to
$2Ge$ where $G$ is the electron lattice coupling constant \cite{13}. The
last term refers to that static elastic energy due to strain
with $`c$' being the relevant elastic constant. Therefore,
because of the electron-lattice ligand potential due to
distortion the original degenerate bands with energy
$(\epsilon_k-\mu)$ now shifts to different subbands
$\varepsilon_{k,\alpha}$ =$\epsilon_k -(-1)^\alpha Ge - \mu$. We
assume that some pairing interaction exists only within the same
orbital and that the strength of the pairing interaction is same
for both the orbitals. Such pairing interaction may be written
as, 
\begin{equation}
H_p = \sum_{k,k^\prime,\alpha,\alpha^\prime} V_{k,k^\prime,\a
\alpha^\prime}C_{k\a\uparrow}^\dagger C_{-k\a\downarrow}^\dagger
C_{-k^\prime\a^\prime\downarrow} C_{k^\prime\a^\prime\uparrow}
\end{equation}
with $V_{k,k^\prime \alpha \alpha^\prime} = -V
\delta_{\alpha,\alpha^\prime}$ and exists only within a cut-off
energy $\pm \omega_D$ around the Fermi level. So, within the
mean field approximation, the model Hamiltonian and
superconductivity may be obtained from equations (1) and (2) as,
\begin{equation}
H = \sum_{k,\alpha \sigma} \varepsilon_{k \alpha} n_{k\alpha \s}
+ \sum_{k,\a}(\D C_{k\a\uparrow}^\dagger
C_{-k\a\downarrow}^\dagger + \D^\star
C_{-k\a\downarrow}C_{k\a\uparrow}) + \frac{\mid\D\mid^2}{V}
+\frac{1}{2} c e^2
\end{equation}
The superconducting order parameter $\D = - \displaystyle \sum_k$
$V<C_{k\uparrow}^\dagger C_{-k\downarrow}^\dagger>$ satisfies the
equation, 
\begin{equation}
1 = \sum_{k,\alpha}\frac{V}{2\omega_{k\a}}
\tanh(\frac{\beta \omega_{k\a}}{2})
\end{equation}
where $\omega_{k\a} = \pm \sqrt{\varepsilon_{k,\a}^2 +\D^2}$ are
the superconducting quasi particle energies in different subbands. The
equilibrium value of strain $e$ is one that minimizes the free
energy and it is given by,
\begin{equation}
e = -\frac{G}{c} \sum_{k,\a} (-1)^\a
\frac{\varepsilon_{k\a}}{2\omega_{k\a}} \tanh(\frac{\beta\omega_{k\a}}{2})
\end{equation}

 Finally, the chemical potential is determined from the
equation for number (n) of charge carriers and hence the hole
concentration as,
\begin{equation}
\delta = \frac{1}{N} \sum_{k,\a}
\frac{\varepsilon_{k\a}}{2\omega_{k\a}}
\tanh(\frac{\beta\omega_{k\a}}{2}) 
\end{equation}
The equations (4-6) are coupled integral equations of order parameters
 $\D$, $e$ and $\delta$. In order to understand the
nature of the coexistence phase, these three equations are
solved self-consistently numerically. The temperature dependence
of the chemical potential $\mu (T)$ is important in determining
the strain and the SC gap parameter self-consistently which has been
taken care such that the calculated value of $\delta$ differs
from the given value of $\delta$ at best at the fifth decimal
point. This is carried out by replacing the $k$-summation by an
integral over $\epsilon$ with an appropriate density of states
$\rho(\epsilon)$. To simulate the strong energy dependence of
$\rho(\epsilon)$ around the centre of the band in the quasi two
dimensional system we substitute $\rho(\epsilon)$ = $\rho(0)
\sqrt{1 - \mid \frac{\epsilon}{B}\mid}\ln
\mid\frac{B^2}{\epsilon^2}\mid$ with $\rho(0) = (\int_{-B}^{B}
\rho(\epsilon) d\epsilon)^{-1}$. Also the attractive potential
$V$ (that appears in equation (2)) is simulated by,
$V(\epsilon)$ = $V_0 [1 -
\frac{(\epsilon-\epsilon_F)^4}{\omega_{D}^4}]^{1/2}$ , in which
$\epsilon_F$ denotes the Fermi energy and $V_0$ the strength of
the pairing potential.
\section{Results and discussions}

 The coexistence of the BJT and the superconductivity is studied mainly by 
self-consistently solving the above set of coupled equations (4 -6) for 
certain values of the input parameters (G, $\omega_D$, c and B) with varrying 
hole concentration $\delta$. The results presented below are derived with the 
dimensionless pairing potential $\rho(0)V_0$ = .11, the cut-off frequency 
$\omega_D = 0.038$ eV, the elastic stiffness constant c = 30 eV and the half-band 
width B=.1 eV. In this section, we first present the results that are obtained
within our model and then discuss its close resemblence with the experimental 
observation later.

The thermal variation of the chemical potential at best small, is found to
be important in determining phase diagram. Fig. 1 depicts the thermal 
variation of strain $e$ for different hole concentrations ($\delta$) 
(indicated in the figure) with a fixed value of electron-lattice coupling 
G (=1.16 eV). It is seen that when the system is closed to half-filling 
(small $\delta$), the structural transition takes place at a quite high 
transition temperature ($T_s$). The strain grows with lowering in 
temperature and would have reached its maximum value at $T =0$ if the
system had remained in normal state (the curve with squares for non-superconducting
(Non-SC) case). On the other hand, very different 
behaviour of $e$ is found when superconductivity occurs at lower 
temperature. The maximum value of strain is reached at the superconducting
transition temperature $T_c$ for all doping and the maximum value decreases
sharply with the increase in hole doping. The cusp -like behaviour of 
$e (T)$ at $T_c$ has important bearing in thermal expansion coefficient 
discussed later. For small $\d$, $e$ saturates at low T ($<<~T_c$), and the
saturation value at T = 0 decreases as dopant concentration goes up.   
Beyond a concentration which depends on model parameter strain disappears
at lower temperature marking a re-entrant behaviour. This means that the
system with higher value of $\d$ goes from undistorted to distorted to 
undistorted phase transformation as temperature is lowered. The extent of
suppression of strain depends sensitively on $\d$ and the electron-lattice
coupling strength G. The structural transition temperature $T_s$ goes down
with $\d$, and for $\d = 0.3$ distortion does not appear.
\par The thermal variation of the superconducting order parameter $\D$ is
presented in Fig. 2. To exemplify the effect of distortion on gap parameter
the variation of $\D$ for $\d = 0.1$ and G = 0 is also shown. The gap 
parameter is much reduced in presence of strain. It is also clear that $T_c$
is suppressed. For $\d = 0.15$ the strain vanishes at lower T ($<T_c$) 
leading to higher value of $\D$. The decrease in $\D$ at low T (T $<<T_c$) 
leading to higher dopant concentration is related to the decrease in the 
density of states at Fermi level in normal state. For reentrant situation a 
discontinuous growth of $\D$ is observed around a temperature where 
strain vanishes. It is also found that the BCS characteristic ratio is not
constant and is higher than 3.5.
%\\
%
%\input{kgp1} \\
%
%\input{phica4} \\
%
\par In Fig. 3(a) the hole concentration dependence of the strain at low
T (T $<< T_c$) for three different values of G is displayed. 
The magnitude of strain is 
highest for zero-doping system in which the Fermi level lies at maximum of
the density of states. At first the strain decreases slowly for low 
$\d$-regime. For G=1.16 eV it vanishes shatply at a critical value of 
$\delta$ whereas in case of G = 1.18 $\&$ 1.2, it drops to
zero discontinuously but at a higher value of $\d$. The strain 
strongly depends on G and increases monotonically with G (Fig. 3b). Its value 
is nearly double for $\d =0.07$ when G is increased by 4$\%$ only. Similarly,
$T_c$ exhibits strong dependence on G (Fig. 3 (c)) and decreases with G. The
extent of decrease is larger for smaller value of $\d$
%\\
%
%\input{phica2.tex} \\
%
\par The dependence of superconducting transition temperature $T_c$ on hole 
concentration is presented in Fig. 4(a). In absence of BJT effect (G = 0 curve)
$T_c$ decreases slowly with $\d$. However, $T_c$ is much lowered in presence
of distortion. The maximum suppression occurs at $\d = 0$ and the extent of
suppression is strong function of coupling strength G. It means that 
whereas the distorted-superconducting (D-S) phase exists for moderate value 
of G, the distorted-normal (D-N) phase can persist down to $T =0$ with 
slightly higher value of G. Another important feature of this result is that
the optimum doping concentration at which maximum $T_c$ is obtainable is
different for different G. An interesting correlation between $T_c$ and
strain at $T_c$ i.e e($T_c$) is observed and is depicted in Fig. 4(b). These
quantities are obtained with different $\d$ nearly collapse into a single
curve over a range of $e(T_c$). This correlation emphasizes the 
role of lattice distortion in determining $T_c$. Incidently, it is 
tempting to note that the such correlation namely lower $T_c$ in more 
distorted system has been observed in $La_{2-x}M_xCuO_4$ (M =Sr, Ca) compound
\cite{10}.
%\\
%
%\input{kgp3}
%\\  
%
\par The thermal expansion coefficient ($\a$) in this model system is much
influenced by the interplay of strain and superconductivity. Apart from the 
normal contribution from the anharmonicity, the thermal variation of strain 
will contribute to $\a$. The thermal coefficient can be expressed as $\a$=
$\a_{anh}$+$\frac{de}{dT}$ where $\a_{anh}\approx \gamma \frac{C_v}{3B}
\cite{12}$ ; $\gamma$ is the Gr\"{u}neison parameter, $C_v$, the specific heat
due to phonon and B, the bulk modulus. The first derivative of $e$ as obtained
from the Fig. 1 for $\delta=0.07$ and 
is shown in  Fig. 5. At low temperature
$(T<<T_c)$  $e$ is suppressed and is nearly independent of
temperature (cf. Fig. 1) so $\frac{de}{dT}$ is nearly equal to 
zero. Therefore, at $T_c$, $\frac{de}{dT}$ has a discontinuity and the magnitude
of discontinuity depends on $\d$.
Above $T_c$ it is negative and very large near $T_s$ and becomes zero again
above $T_s$. This will produce a positive discontinuity at $T_c$ and a 
minimum at $T_s$ in the thermal variation of $\a$. The minimum is expected 
to be broadened in presence of fluctuation of strain parameter. This result
is in qualitative agreement with experimental observation in oxide 
superconductors \cite{8}.
\par Finally, the phase diagram $(T_s, T_c~vs~\delta$) is drawn in Fig. 6 
for two close values of G. It shows
the evolution of the distorted-normal (D-N), coexistence of
distorted superconducting (D-S), undistorted superconducting (UD-S)
 phases as well as
the re-entrance of the undistorted phase with hole concentartion.
In Fig. 6, the bold dots (solid line) represent the structural (superconducting)
line indicating the respective phase boundaries. 
The transition temperature $T_s$, where distortion appears is highest
for zero-hole concentration $\d =0$ (half-filled case). On the other hand,
the superconducting transition temperature $T_c$ which is less than 
$T_s$ attains its maximum value at certain concentration. Below that concentration
$T_c$ is much suppressed compared to 
its maximum value and is also suppressed compared to its value for 
undistorted phase. With the increase of $\d$, $T_s$ decreases whereas
$T_c$ goes up. The two transition temperatures become equal at a critical
concentration of $\d_c$ which depends on the values of the input parameters.
Above $\d_c$, the structural transition is completely inhibited by the 
onset of superconductivity and $T_c$ decreases again. Therefore, the 
highest value of $T_c$ is determined by structural transition temperature
$T_s$ in accordence with the experimental observation \cite{8}. 
Below $\d_c$ a small region in $\d$ characterized by
vanishing of $e$ at low temperature is found. At $ T= 20$K, the distorted phase
appears at a concentration $\d \leq 0.15$ which is less than $\d_c$. 
Therefore, within this region of hole doping the phase transition from D-N to
D-S to UD-S is observed with lowering of temperature. The re-entrant 
region shrinks with increasing value of G. In summary, the salient features of
the phase diagram as follows from the model are : \\
\noindent (i) The spontaneous distortion appearing first at higher temperature 
interferes destructively with superconducting tendency of the system. However,
the superconductivity co-exists with distortion within a certain region of parameter 
space. \\
\noindent (ii) With slight increase in electron-lattice interaction strength 
(G) the 
superconductivity may be inhibited in a system with low doping 
(cf. Fig. 4(a) also)\\
\noindent (iii) a reentrant structural transition where system passes from
undistorted normal phase to undistorted superconducting one through 
intervening distorted normal and distorted superconducting 
phases exist within small range of $\d$ (cf. Fig. 1 also). \\
\noindent (iv) The maximum value of $T_c$ is limited by $T_s$ and $\d_c$ \\
\noindent (v) The distorted phase does not appear if 
superconductivity occurs first. 

\par The structural transition in this model results from the 
competition between lowering of the electronic energy due to lifting of the
orbital degeneracy (band Jahn-Teller effect) and the increase in elastic
energy associated with the strain. The lowering of electronic energy 
is associated with the redistribution of electronic states due to removal
of orbital degeneracy. The net gain in energy on this process depends on
the location of the Fermi level in the undistorted phase. When the lowering
of the electronic energy due to  distortion overwhelmes the increase in 
strain energy, the spontaneous distortion appears in the system. The measure
of this gain in energy is the structural transition $T_s$ and this gain 
becomes largest when the Fermi level in undistorted phase lies at the maximum
of the density of states. Both the transition temperature $T_s$ and the 
strain $e$ (T = 0) are maximum for $\d =0$ as the Fermi level lies at 
singular point of density of states. With the increase in hole concentration
the Fermi level moves away from point of large density of states and this 
leads to lowering in $T_s$ and $e$. The suppression of the 
superconducting transition temperature in presence of structural distortion 
is related to removal of states from the Fermi level due to lifting 
of degeneracy. Larger the splitting of orbital states (higher values of 
$T_s$ and $e$) larger is suppressive effect on $T_c$. We note that 
the superconductivity can appear in system with or without distortion 
depending on electron-lattice coupling. However, it is clear that the highest
value of $T_c$ can be achieved in undistorted phase. With increase of hole
concentration the suppressive effect (removal of states at the Fermi level) 
is diminished with conseqent increase in $T_c$. In distorted-superconducting 
phase there is destructive interference between superconducting order parameter
$\D$ and strain $e$. The lowering of strain in superconducting phase has to do
with the appearence of energy gap accross the Fermi level. The existence of 
the gap reverses the trend of lowering of energy due to structural distortion. 
The distortion is compeletly suppressed when the superconducting gap is larger
than the orbital splitting energy (2$Ge$). The reentrant behaviour at 
low temperature is the result of mutual influences between two order 
parameters. It is observed that the nature of the density of states and the 
location of the Fermi level are important ingredients in determing competing
aspects of the superconductivity and the BJT like distortion. The maximum 
competing effect between the two order parameters can occur in the system with 
Fermi level located at energy where the density of states is large and 
sharply varying.
\par We would like to note that many salient features of the results 
are in tune with the trend of the experimental observations. The interplay 
between superconductivity and distortion has been distinctly observed in
$La_{2-x}M_xCuO_4$ (M =Sr, Ca, Nd) \cite{10}. The distortion at very low 
temperature decreases with increase in $x$ (hole concentration) and vanishes 
at a critical value of $x$. The $Ca-$substituted system is slightly more 
distorted than that for $Sr$- substituted one and at the critical value of
$x$, the distortion vanishes with slight discontinuity in former compound in
contrast to smooth vanishing for later one. For a fixed $Ca$ concentration, the 
distortion increases with $Nd$ concentration which replaces $La$. These results are similar
in nature to those in Fig. 3(a,b) $\&$ 3(c), if it is assumed that the
$Ca$-substituted compound has higher value of the electron-lattice coupling 
constant than that of $Sr$-substituted system. Similarly assuming 
$Nd$- substitution augments the value of G, the results of Fig. 3(a,b) simulates 
the observation that strain ($T_s$) increases (decreases with 
$Nd$ concentration). The higher value of G in $Ca$ ($Nd$) - substituted 
compound is related to lower value of ionic size of $Ca$ ($Nd$) which leads 
to smaller $La(Ca)-O$ $(La(-Nd)-O)$ bonds. The smaller bond length in 
turn increases the ligand field seen by hole. As G is determined by 
the rate of change of ligand potential one expects larger value of G for $Ca
(Nd)$ - compound in comparison to $Sr$-based compound. This assumption 
is consistent with experimental observation that the lower $T_c$ is found in 
more distorted $Ca$-system and $T_c$ decreases with $Nd$-substitution for 
doped ($La-Nd$) system \cite{10}. The result that the increase in distortion
lowers $T_c$ as observed in these systems \cite{10,17} is in harmony with 
depicted in Fig. 3. Moreover, general features of the phase diagram
(Fig. 5) are close to observed phase diagram in these system \cite{10}. The
fact that the highest $T_c$ is limited by $T_s$, the coexistence of
superconductivity and distortion and complete suppression of superconductivity
in system with large distortion are borne out from the model. For $G =116$,
$T_s$ varies little with $\d$ and the maximum $T_c$ is $T_s$ at $\d_c$. This
correlates well with the observation in cuprates \cite{8}. As noted earlier
\cite{10} the crucial quantity in the interplay between the two 
order parameters is the density of states at FL. The distortion produces a
notch at FL which in turn lowers $T_c$. The notch gets deeper for 
larger distortion and so the larger suppression of $T_c$ results. As the FL 
shifts from singular point with doping the redistribution of DOS around 
FL is less marked and hence the suppressive effect of distortion on $T_c$
is lessened.
\section{Conclusion}
In conclusion, a simple model based on orbitally degenerate band
with electron-lattice coupling and the BCS-type pairing demonstrates
the interplay between structural transition and superconductivity. It
is found that the order parameters interferes destructively. The
suppression of $T_c$ in the distorted phase is due to small number of
states available at the Fermi level for pairing. The states at the Fermi
level is depleted due to lifting of orbital degeneracy. This effect is
highest when the Fermi level lies at the peak of the density of states. 
However, the distortion and the superconductivity can coexist in 
system for limited range of parameters of the model. The higher 
value of $T_c$ is found in system with lower value of distortion. 
With increase of hole concentartion $T_c$ increases and the 
distortion is completely removed at a critical hole concentration.
The value of $T_c$ at this concentration where $T_s = T_c$ is its 
highest value. The existence of the superconducting state at lower 
temperatures depend very sensitively on the strength of 
electron-lattice coupling. The effect of ionic size of iso-electronic $M$ 
$(Sr, Ca, Nd$) ion on the interplay of superconductivity and structural
transition can be understood as the consequence of higher electron-lattice
coupling. The electronic correlation effect are not discussed here and will 
reported elasewhere \cite{16}.

\end{document}